
\font\titlefont = cmr10 scaled \magstep2
\magnification=\magstep1
\vsize=20truecm
\voffset=1.75truecm
\hsize=14truecm
\hoffset=1.75truecm
\baselineskip=20pt

\settabs 18 \columns

\def\b{\bigskip}
\def\bb{\bigskip\bigskip}
\def\bbb{\bigskip\bigskip\bigskip}

\def\ce{\centerline}

\def\no{\noindent}




 \rightline{ UMDHEP 94-30}
\rightline{ August 1993}
\bb

\b
\ce{\titlefont{Lepton Flavor Violation and the Tau Neutrino Mass }}

\bb
\ce{\bf{R.N. Mohapatra\footnote\dag{\rm{ Work supported by a grant from the
National
          Science Foundation}},~~
S. Nussinov\footnote{*}{\rm{Permanent address: Department of Physics and
Astronomy, Tel Aviv University, Tel Aviv, Israel.}}
  and X. Zhang{$\dag$}
   }}

\ce{\it{ Department of Physics and Astronomy}}
\ce{\it{University of Maryland}}
\ce{\it{ College Park, MD 20742 }}
\b
\ce{\bf Abstract}

\no
We point out that,
in the left-right symmetric model of weak interactions,
 if $\nu_\tau$ mass is in the keV to MeV range,
 there is a strong
correlation between rare decays such as
$\tau \rightarrow 3 \mu, ~~\tau \rightarrow 3 e$ and
the $\nu_\tau$ mass. In particular, we point out that a large range of
$\nu_\tau$ masses are forbidden by
the cosmological constraints on $m_{\nu_\tau}$ in combination with
 the present upper
limits on these processes.

 \filbreak
In the standard model of electroweak interactions, all lepton flavors
$L_e, ~~L_\mu ~~{\rm and}~~L_\tau$ are conserved. On the other hand, in most
extensions of the standard model, lepton flavor conservation is not maintained;
therefore, it is hoped that the nature of lepton flavor violation can help to
narrow the possibilities of new physics.
Crucial tests of lepton flavor violation are provided by the
rare decays of $\mu$ and
$\tau$ such as $\mu \rightarrow 3 e$[1] and
$\tau \rightarrow l_i l_j {\overline l_k}$ where
$l_{i,~j,~k}$ go over e
and $\mu$.
The present stringent limits on $\mu \rightarrow 3 e$
already make it imperative that in all extensions of standard model,
violation of $L_\mu ~+~ L_e$ be very weak. On the other hand, the
present upper limits on the branching ratios for
rare $\tau$ decays[2]
allow for possible violation of $L_\tau ~+~ L_\mu$
or
$L_\tau ~+~ L_e$ at
a much higher level.
One class of models, where the possibility of a significant lepton violation
exists, is the left-right symmetric model with
see-saw mechanism for neutrino masses[3].
 In this note, we investigate the rare $\tau$ decays
and their implications for violation of $L_\tau ~+~ L_\mu$
or $L_\tau ~+~ L_e$ quantum numbers in these models.
 We show that
there is a strong correlation
between the tau neutrino mass and $\tau \rightarrow 3 \mu$
and
$\tau \rightarrow 3 e$ decays
 if $m_{\nu_\tau}$
is in the keV to MeV range[4], as is allowed by
existing laboratory upper limits[2].
First we derive lower limits on $m_{\nu_\tau}$ for the case where
$B( \tau \rightarrow 3 \mu ) = 0$; Once the flavor violating
decay $\tau \rightarrow 3 \mu ~~{\rm or}~~ \tau \rightarrow 3 e$
is allowed we show that
the present upper limits on their rates permit the
lower bound on $m_{\nu_\tau}$ to be
somewhat relaxed.
Improvement of the present experimental upper limits on $m_{\nu_\tau}$
and the branching ratios for the above $\tau$-decay modes can therefore throw
light on the nature of lepton flavor
violation in the left-right symmetric models.

We consider
the left-right symmetric model with
a see-saw mechanism for neutrino masses as
described
in ref.3. Let us display the leptonic and Higgs sector of the model.
The three generations of
 lepton fields are $\Psi_a \equiv {\pmatrix{\nu \cr e \cr }}_a$,
where $a~ =~ 1,~ 2,~ 3$.
 Under the
gauge group $SU(2)_L \times SU(2)_R \times U(1)_{B-L}$, they transform as
$\Psi_{a~L} \equiv (1/2, ~ 0 , ~ -1 )$
and $\Psi_{a~R} \equiv (0, ~ 1/2, ~ -1 )$.
Since our purpose is to study the
possible degree of violation of $L_\mu ~+~ L_\tau$
or $L_e ~+~ L_\tau$ in the rare $\tau$-decay, we will impose
one of
these global
symmetries on the model[F.1], for
simplicity. We illustrate our idea for the model with
$U(1)_{\tau ~+~ \mu}$ global symmetry. The Higgs sector then needs to be
enlarged if we want the see-saw
mechanism for all lepton flavors. We choose a single bi-doublet field
$\phi \equiv (1/2, ~ 1/2, ~ 0)$ and two sets of triplet Higgs fields:
$${
\Delta_L ( 1, ~0, ~ +2 ) \oplus \Delta_R (0, ~ 1, ~ +2 );
{}~~~~~~{\rm with} ~~L_\mu + L_\tau = -2 ~~;}
\eqno(1. a)$$

$${
\Delta^{\prime}_L (1, ~ 0, ~ +2 ) \oplus \Delta_R^{\prime}
                           (0, ~ 1, ~ +2 );
{}~~~~~~ {\rm with}~~ L_\mu + L_\tau = 0 ~~.}
\eqno(1. b)$$

\no The Yukawa coupling
which are invariant under all symmetry can be written as:
$${
\eqalign{
{\cal L}_Y
= & {\overline \Psi_L} h \phi \Psi_R + {\overline \Psi_L} {\tilde h}
 {\tilde \phi} \Psi_R \cr
   & + \Psi^T_L \tau_2 {\vec \tau} \cdot f {\vec \Delta_L} C^{-1} \Psi_L
                                + L\rightarrow R  \cr
               & + h.c     ~~~~.
}}\eqno(2)$$
\no where
$h, ~{\tilde h}$ and
 $f {\tilde \Delta}$ are the following matrices in generation
space:
$${
h \equiv \pmatrix{
h_{11} & 0 & 0\cr
0 & h_{22} & h_{23} \cr
0 & h_{23} & h_{33} \cr}
} \eqno(3.a)$$

$${
f {\vec \Delta} \equiv \pmatrix{
f_{11}{\vec {\Delta^{\prime}} } & 0 & 0 \cr
  0 & f_{22} {\vec \Delta} & f_{23} {\vec \Delta} \cr
  0 & f_{23} {\vec \Delta} & f_{33}{\vec \Delta} \cr }
}\eqno(3.b)$$

\no and similarly for $\tilde h$.

The gauge symmetry is spontaneously broken by the vacuum expectation
values:
$${
< {\Delta_R^0} > = V_R ~~; ~~ < {\Delta_R^0}^{\prime} > = V_R^{\prime}
{}~~; }\eqno(4.a)$$

$${
< \Delta_L^0 > = < {\Delta_L^0}^{\prime} > = 0
}\eqno(4.b)$$
\no and
$${
< \phi > = \pmatrix{ \kappa & 0 \cr
                    0 & \kappa^\prime \cr } ~~.
}\eqno(4.c)$$

\no As usual, $< \phi >$ gives masses to the charged fermions and Dirac masses
to the neutrinos whereas
$< \Delta_R^0 >$
and
$< {\Delta_R^0}^\prime >$ lead to the see-saw mechanism for the neutrinos
( this mechanism
operates separately for $\nu_e$ and
jointly for $\nu_\mu$
and
$\nu_\tau$ ).
These discussions are all standard and we do not
repeat them here.

The physics we are interested in comes from the left-handed triplet
sector of the theory.
As indicated in eq.(4.b)
these fields do not take part in the Higgs mechanism;
therefore, if we ignore certain
couplings in the Higgs potential, such as $\Delta_L \phi \Delta_R^{\dagger}
  \phi^{\dagger}$ etc, then
$\Delta_L ~~{\rm and}~~ \Delta_R$
remain unmixed states. Of course there could be mixings between $\Delta_L$
and $\Delta_L^\prime$; but we ignore these
mixings here and comment
later on their effect.
For small
$\Delta_L - \Delta_L^\prime$
mixings, our main results
do not change. In this limit, the electron generation separates for all
 practical
purposes from the $\mu$ and
$\tau$ generations.
The
$\mu \rightarrow 3 e$
and
$\mu \rightarrow e \gamma$ are forbidden. We therefore focus on the
$\mu - \tau$ sector. The Yukawa Lagrangian relevant
to our discussions is given ( in the basis where all the leptons are mass
 eigenstates ) by

$${
\eqalign{
{\cal L}_Y = & \nu_L^{T} F^{\prime} C^{-1} \nu_L \Delta^{0}
                  + \nu_L^T F^{''}C^{-1}E_L \Delta^{+}_L \cr
             & + E_L^T F C^{-1} E_L \Delta^{++} + h.c ~~~~~~,
                 }
}\eqno(5)$$

\no where
$\nu = ( \nu_\mu, ~ \nu_\tau ),~ E = ( \mu , ~ \tau ); ~F, ~ F^\prime ~
{\rm and} ~ F^{''}$ are
$2\times 2$ matrices related to each other
as follows:
$${
F K^T = F^{''} ; ~~ K F K^T = F^\prime ~~~,
}\eqno(6)$$
\no where $K$ is the leptonic Cabibbo
 matrix in the left-handed
$\mu - \tau$ sector. First
we note that the off-diagonal element of
$K$ is the
$\nu_\mu - \nu_\tau$ mixing angle, which
is directly
measurable parameters, restricted to be,
 $\theta_{\mu\tau} \leq 0.03$
  by existing
accelarator experiments[5].

Now, we make the following observation.
Suppose that the $\nu_\tau$ mass is in the keV to MeV range and
$\nu_\mu$
and $\nu_e$ masses are in the few electron volt range, a possibility consistent
with present upper limit on neutrino masses from the accelerator data[5].
In this case, $\nu_\tau$ must be unstable in order to be consistent with
cosmological constraints on the mass density in the universe[6].
The mass and life time are then related by[7]

$${
\tau_{\nu_\tau}
\leq ( 5.4 \times 10^{10} {\it sec}) { \left(
 { 100~ {\rm keV} \over m_{\nu_\tau} } \right) }^{2}
}\eqno(7.A)$$

\no A
 more stringent,
but model dependent
 constraint can be derived from considerations of Galaxy
formation[8]; it is given by

$${
\tau_{\nu_\tau} \leq 3 \times 10^7 {\it sec}.
}\eqno(7.B)$$

\no In the model under consideration, $\nu_\tau \rightarrow
  \nu_\mu ~ {\overline \nu_\mu}~ \nu_\mu$ occurs via
$\Delta^0_L$
exchange and can be used to satisfy the constraints in eqs.(7)[9]. The
 Hamiltonian for this process is given by

$${
H = {G_{\nu_\tau} \over \sqrt2}
    {\overline \nu_\mu} \gamma^{\lambda}
          ( 1- \gamma_5 ) \nu_\mu
         {\overline \nu_\mu} \gamma_{\lambda}
             ( 1- \gamma_5 ) \nu_\tau + h.c ~~~,
}\eqno(8.a)$$

\no where (we drop the subscript L from
$\Delta_L$ henceforth)

$${
G_{\nu_\tau} = {\sqrt 2}{ F_{\mu\mu}^{\prime}
                         F_{\mu\tau}^{\prime}\over
                  { 4 M_{\Delta^0}^2 } }
\simeq {\sqrt 2 \over {4 M_{\Delta^0}^2 } }
                     F_{\mu\mu} \times [
                F_{\mu\tau} - \theta_{\mu\tau} (
             F_{\mu\mu} - F_{\tau\tau} ) ]~~~.
}\eqno(8.b)$$

\no The $\nu_\tau$ lifetime is

$${
{\tau_{\nu_\tau}}^{-1}
 = { 2 G^2_{\nu_\tau} m_{\nu_\tau}^5 \over {192 \pi^3} }
{}~~~~.}\eqno(8.c)$$

\no From eq.(7.A) we get[F.2]
$${
G_{\nu_\tau} \geq (1.9 \times 10^{-12}) { ( { {\rm GeV} \over
             m_{\nu_\tau} }) }^{3/2}  {\rm GeV}^{-2}
                            ~~~~. }
\eqno(9.A)$$

\no To get a feeling for the order of magnitude of $G_{\nu_\tau}$, note that
for $m_{\nu_\tau} = 10$ MeV,
$G_{\nu_\tau} \geq 2 \times 10^{-4} G_F$
and
$m_{\nu_\tau}= 0.1$ MeV,
$G_{\nu_\tau} \geq 0.2~ G_F$.
 The corresponding constraint from Galaxy formation
eq.(7.B)
can be written as
$${
G_{\nu_\tau} \geq 8 \times 10^{-15} { ( {  {\rm GeV} \over m_{\nu_\tau}} )
                               }^{5/2}~{ \rm GeV}^{-2} ~~~~~~.
}\eqno(9.B)$$

Turning now to the $\tau$-lepton, we observe that exchange of
$\Delta_L^{++}$ contributes to the rare $\tau$-decay,
$\tau^{-} \rightarrow
\mu^{-} \mu^{-} \mu^{+}$ with a strength (defined analogously to
     the $\nu_\tau$ case )
$${
G_\tau = {\sqrt 2}
{ F_{\mu\mu} F_{\mu\tau} \over { 4 M_{\Delta^{++}}^2 } }
 ~~~~~~~. }
\eqno(10)$$

\no Now, we first notice from eq.(8.b) that, even if $F_{\mu\tau}=0$
( {\it i.e} there is no $\tau \rightarrow 3 \mu$ decay ) the
$\nu_\tau$ can decay. Since the decay rate depends on $\nu_\tau$ mass,
let us see, if for the presently allowed range of
$\theta_{\mu\tau}$ and
$\nu_\tau$
masses, constraints in eqs.(7.A) and (7.B) are satisfied.
To study this, we first note that vacuum stability requires all
$F_{a b} \leq 1.2$[11]
and LEP
data require that, $m_{\Delta_L^{0}} \geq 45$ GeV. Combining these and
present upper limit of $\theta_{\mu\tau} \leq 3 \times 10^{-2}$
we find from eqs.(9) that, for case A and B, the $\nu_\tau$
mass must have the following lowest bounds:
$${
{\rm Case~ A:}
{}~~~~~~m_{\nu_\tau} \geq 31~ {\rm keV} ~~~~~~;
}\eqno(11.A)$$

$${
{\rm Case~ B:}
{}~~~~~~ m_{\nu_\tau} \geq 210~ {\rm keV}~~~~~~.
}\eqno(11.B)$$

\no Once the $\nu_\tau$ masses go below the above limits, the LR model cannot
satisfy
the cosmological constraints without having $F_{\mu\tau} \not= 0$.
We emphasize that, we have been extremely conservative in obtaining
these lower bounds. (For instance,
$F_{\tau\tau}$ is likely to be lower than its maximum allowed value and
$m_{\Delta^0}$ is also likely to be heavier.)
  $F_{\mu\tau} \not= 0$ immediately leads to non-vanishing
$\tau \rightarrow 3 ~\mu$ decay.
  We can therefore obtain a lower bound on the
$B( \tau \rightarrow 3 \mu )$ in these ranges of $m_{\nu_\tau}$.

In the presence of $F_{\mu\tau}$, we have

$${
\Gamma
(\tau \rightarrow 3 \mu )
= {1 \over {4 m_{\Delta^0}^4 }}
               { ( F_{\mu\mu}F_{\mu\tau} )}^{2}
               {m_{\tau}^5 \over {192 \pi^3 }} ~~~~~~.
}\eqno(12)$$

\no Using the $\nu_\tau$ lifetime
in eq.(8.c), we get,

$${
B( \tau \rightarrow 3 \mu )
 = {({ 0.3 \times 10^{-12} \over {\tau_{\nu_\tau}~{\rm in {\it sec}} }})}
                    {( {m_{\Delta^0} \over m_{\Delta^{++}} })}^{4}
                    {( { m_\tau \over m_{\nu_\tau} })}^5
       { ( { F_{\mu\tau} \over {
             F_{\mu\tau}- \theta_{\mu\tau} (F_{\mu\mu} - F_{\tau\tau}) } })}^2
                 .
}\eqno(13)$$

\no Using the cosmological upper bounds on $\tau_{\nu_\tau}$ in eqs.(7.A) and
(7.B),
we get,

$${
{\rm Case ~A:}~~
B( \tau \rightarrow 3 \mu ) \geq 9.5 \times 10^{-3}
                    {( {100 {\rm keV} \over m_{\nu_\tau} })}^3
                   {( { m_{\Delta^0} \over m_{\Delta^{++} } })}^4
                   {\epsilon}_{\mu\tau} ~~~;
}\eqno(14.A)$$

$${
{\rm Case ~B:}~~~~~~
B( \tau \rightarrow 3 \mu ) \geq 16.8
   {( { 100 {\rm keV} \over m_{\nu_\tau} })}^5 {( {
    m_{\Delta^0} \over m_{\Delta^{++} } })}^4 {\epsilon}_{\mu\tau}
 ~~~.
}\eqno(14.B)$$

\no where $\epsilon_{\mu\tau} = {( {
F_{\mu\tau} \over {F_{\mu\tau} - \theta_{\mu\tau}(F_{\mu\mu}-F_{\tau\tau})
} })}^2 $.
 Note that once $m_{\nu_\tau}$
is below the lower bounds given in eqs.(11),
$\epsilon_{\mu\tau}$ becomes a function of $m_{\nu_\tau}$;
therefore for a given value of $m_{\nu_\tau}$, we can find a lower bound on
$B( \tau \rightarrow 3 \mu )$, (and vice-versa) if we have a lower bound on
${( { m_{\Delta^0} / m_{\Delta^{++} } })}^4 $.

Let us
therefore discuss the factor ${ (m_{\Delta^0} / m_{\Delta^{++} } )}^4$.
We note that[10], the
$\vec {\Delta_L}$ multiplet contributes to the $\rho$-parameter as follows:
$${
\rho_\Delta
 = { G_F \over { 4{\sqrt 2} \pi^2 } }[ f_{(0, +)}
   + f_{(+, ++)} ] \equiv { 3 G_F \over {8 {\sqrt 2} \pi^2 } }
  {\Delta m^2}
   ~~~, }
\eqno(15)$$
\no where
$f_{a, b}
= m_a^2 + m_b^2 -
{2 m_a^2 m_b^2 \over {m_b^2 - m_a^2 } } \ln {m_b^2 \over m_a^2}$.
Langacker[12] has given an upper bound on the new contribution
to $\rho$-parameter from physics beyond the standard model as follows:
$${
m_t^2 + \Delta m^2 \leq {( {194 {\rm GeV} })}^2
{}~~~~.}\eqno(16)$$

\no In the LR model, there exists the further relation:
$${
m_{\Delta^{++}}^2 = m_{\Delta^0}^2 ( 1 + 2 \alpha ); ~~~
m_{\Delta^+}^2 = m_{\Delta^0}^2 ( 1 + \alpha ) ~~,
}\eqno(17)$$

\no where $\alpha$ is a dimensionless parameter.
 Using these relations, we can obtain a lower bound on $\alpha$ from
$ \rho$-parameter
 constraint
(using the fact
that $m_{\Delta^0} \geq 45$ GeV),
 which can then be converted to a lower bound on
$B( \tau \rightarrow 3 \mu )$.
We find that for $m_t = 110 ~{\rm GeV}~, ~ \alpha < 67 ~~{\rm and ~ for}~
m_t = 150, ~ \alpha < 40$. Using this we obtain

$${
{\rm Case~ A:}~~~
 B( \tau \rightarrow 3 \mu ) \geq \delta_A ~{( {
                            100 {\rm keV} \over m_{\nu_\tau} })}^3
                           \epsilon_{\mu\tau}( m_{\nu_\tau} )
{}~~;
}\eqno(18.A)$$

$${
{\rm Case~ B:}~~~
B( \tau \rightarrow 3 \mu )
   \geq \delta_B~ {({ 100{\rm keV} \over m_{\nu_\tau}})}^5
               \epsilon_{\mu\tau}(
                                m_{\nu_\tau})
{}~~.
}\eqno(18.B)$$

\no In table I, we give the values of $\delta_A ~{\rm and}~
\delta_B$ for the two cases for two values of $m_t$.

\bbb
\b
\bbb
\b
\vbox{\tabskip=0pt \offinterlineskip
\def\tablerule{\noalign{\hrule}}
\halign to300pt {\strut#& \vrule#\tabskip=1em plus2em&
  \hfil#& \vrule#& \hfil#\hfil& \vrule#&
    \hfil#& \vrule# \tabskip=0pt \cr \tablerule

&& $m_t$ {\rm GeV}  &&  $\delta_A$ && $\delta_B$
                                                               &\cr \tablerule
&& 110  &&  $5.6\times 10^{-7}$  &&  $1 \times 10^{-3}$   &\cr \tablerule
&& 150 && $1.4 \times 10^{-6}$ && $2.5 \times 10^{-3}$ &\cr \tablerule
 \noalign{\smallskip}
& \multispan{7} Table I. Values of $\delta_A~ {\rm and} ~\delta_B$
\hfil \cr }}

\bbb
\bb
\b
\bbb
\bb
\filbreak
\bb
\b

To
understand
the implications of eqs.(18) further, let us first note that they depend on
$F_{\mu\tau}$ explicitly. Clearly for values of $m_{\nu_\tau}$ far below the
lower limits in eqs.(11), cosmology would require $F_{\mu\tau} \gg
\theta_{\mu\tau}( F_{\mu\mu} - F_{\tau\tau}) $
(${\it e.g.} ~~ m_{\nu_\tau}= 100 ~keV$
in case B would require $F_{\mu\tau} \simeq 0.2$,
whereas \hfil\break
${ | \theta_{\mu\tau}( F_{\mu\mu}-F_{\tau\tau} ) | }_{max} \leq .06$ ).
In such cases, $\epsilon_{\mu\tau}=1$ so that, the lower bound is obtained
by setting $\epsilon_{\mu\tau}=1$ in the right-hand side of the inequalities
(18.a) and (18.b).
The present upper bound on
$B( \tau \rightarrow 3 \mu ) \leq 4.8 \times 10^{-6}$[13].
Therefore, values of $m_{\nu_\tau}$ for which
$F_{\mu\tau} \leq {1\over 3} {|\theta_{\mu\tau} ( F_{\mu\mu} - F_{\tau\tau} )
| }_{max}
\simeq .02$ satisfies both the experimental upper bound on
$B( \tau \rightarrow 3 \mu )$ and
the cosmological bound for case A ( $m_t = 150$ GeV ) leading to
$m_{\nu_\tau} \geq 26$ keV, which is, then, the absolute lower bound on
$m_{\nu_\tau}$
in this
model. Turning to case B, we find that both constraints
are satisfied for $F_{\mu\tau} \leq .02$ giving
$m_{\nu_\tau} \geq 187$ keV. Thus, we see that allowing for
$\tau \rightarrow 3 \mu$ decay leads to slight relaxation
of the lower bounds on
$m_{\nu_\tau}$
allowed in the LR model. Further improvement of the upper limits
on the
$B( \tau \rightarrow 3 \mu )$ as well as
$\theta_{\mu\tau}$ will therefore help to further constrain
the $m_{\nu_\tau}$ in these models.

Bounds on $m_{\nu_\tau}$ using only cosmological mass density constraints
were discussed
 in ref.4, where two assumptions were made: a)
$m_{\Delta^0} \simeq m_{\Delta^{++}}$ and
b) there is no mixing between lepton generations. We ${\bf do ~~not}$
make these assumptions here; further more, we
point out the existence of a bound even if
$B( \tau \rightarrow 3 l) = 0$ unlike that in ref.4. Thus, our
lower bounds are more rigorous than those of ref.4.

Let us close with a few comments:

\item{a)} For the case where $L_e + L_\tau$ symmetry is imposed on the theory,
similar results follow with $\mu$ replaced by e everywhere in the final
state. The lower bounds are now weaker since
$\theta_{e \tau} \leq .17$. All result obtained for case A (${\it i.e.}$
mass density bound ) are lowered by a factor of 3 and for case B
by a factor of 2. If
$m_{\nu_\tau} > 1$ MeV, the channel $\nu_\tau \rightarrow
\nu_e ~+~ e^+ ~+~ e^-$ can also arise via $\Delta_L^+$
exchange, which is constrained by Supernova consideration[14],
although the existing bounds do not yield any interesting
constraint on the parameters under discussion.

\item{b)} In the model with $L_\mu + L_\tau$ symmetry, the existence
of
$\Delta_L - \Delta_L^\prime$ mixing
can lead to $L_\mu + L_\tau$ violating channels.
 We choose this mixing to be small.
If however, this mixing were
 not negligible a new channel
${\overline \nu_\tau}
 \rightarrow \nu_\mu ~+~ \nu_e ~+~ \nu_e$ appears. This
 will weaken our bounds by
a factor ${(1 + y)}^{3/2}$ in case A and
${( 1 + y )}^{5/2}$ in case B, where
$y = \Gamma (\nu_\tau \rightarrow \nu_\mu \nu_e \nu_e ) / \Gamma(
\nu_\tau \rightarrow 3 \nu_\mu )$.

\item{c)}
Specifically,
our results
will also apply to the
minimal left-right symmetric model
 without any symmetry (and hence only a single
set
of $\Delta_L \oplus \Delta_R$ )
if we only chose either $F_{\mu\tau}$
or $F_{e \tau}$
to be zero,
except that in this case
there is always a second
decay mode ( ${\it e.g.} ~\nu_\tau \rightarrow
{\overline \nu_\mu} \nu_e \nu_e ~~{\rm for} ~~ F_{e \tau} =0$ ).
Again, there will be a slight dilution of our lower limits.

\item{d)} Strictly speaking in order to avoid
the existence of a Majoron in our model, we can add soft symmetry
breaking terms of the form $( \Delta^{\dagger}_L \Delta_L^\prime
                           + \Delta_R^\dagger \Delta_R^\prime )$.
In the absence of this,
 there exists a Majoron, but it does not provide any fast decay
mode for $\nu_\tau$
(similar to
 the original singlet Majoron model). In either case, our results
 remain unchanged.

\bb
\b
We wish to thank A. Jawahary for discussions
and to Y. Nir for pointing out some earlier works on the subject.
 One of us (S.N.)
thanks the Nuclear theory group at the University of Maryland for hospitality.

\bb
\filbreak

\b
\ce{\bf Footnote}
\item{[F.1]} Even though we have imposed the global symmetries
$L_\tau ~+~ L_\mu$ to obtain these results, these are
very likely to apply to the model without them. The reason is that, $\mu
\rightarrow 3 e$ requires the $\Delta_L$ coupling
$F_{\mu  e}$ to be nearly zero. Then the constraints of $\mu \rightarrow e
\gamma$ imply that, $F_{\tau \mu} F_{\tau e} \leq 10^{-5}$.
Therefore, if one of them is big ({\it i.e} of order $10^{-1}$ or so ),
the second one is very small. In our case, if $F_{\mu \tau } \simeq 10^{-1}$,
we expect $F_{\tau e} \leq 10^{-4}$. This is equivalent to approximate
$L_\mu + L_\tau$ symmetry. Similarly if $F_{\mu\tau} \leq 10^{-4}$
and
$F_{e \tau } \simeq 10^{-1}$, this is equivalent to imposing $L_e ~+~ L_\tau$
symmetry.

\item{[F.2]}Similar
 considerations are applied in the $e-\mu$ sector in ref.10.

\b
\filbreak
\bb

\ce{\bf References}
\b
\item{[1]}A. Van der Schaaf,
Prog. Part. Nucl. Physics, $\bf 31$, 1 (1993);
M. Cooper, Proceedings of LEMS'93, ed. M. Leon (to be published).
\item{[2]}
For an excellent recent review, see A.J. Weinstein and R. Stroynowski,
Cal. Tech. Preprint, CALT-68-1853, Febuary, 1993. to appear in Ann.
Rev. of Nucl. and Particle Physics.

\item{[3]}R.N. Mohapatra and G. Senjanovi\'c , Phys. Rev. Lett.
 $\bf 44$, 912 (1980); Phys. Rev. $\bf D23$, 165 (1981).

\item{[4]}
     H. Harari and Y. Nir, Nucl. Phys. $\bf B292$, 251 (1987).

\item{[5]}F. Boehm, in ${\it Particles,~~ Strings~~ and
{}~~ Cosmology}$, ed. P. Nath ${\it et. al.}$ P.96 (World
Scientific, 1991).

\item{[6]}D. Dicus, E. Kolb and V. Teplitz, Phys. Rev. Lett.
   $\bf 39$, 169 (1977).

\item{[7]} E. Kolb and M. Turner, Phys. Rev. Lett. $\bf 67$, 5 (1991).

\item{[8]} G. Steigman and M. Turner, Nucl. Phys. $\bf B253$, 375 (1985).

\item{[9]}
     M. Roncadelli and G. Senjanovi\'c , Phys. Lett. $\bf B107$, 59 (1983);
      P.B. Pal, Nucl. Phys. $\bf B227$, 237 (1987); R.N. Mohapatra and P.B.
    Pal, Phys. Lett. $\bf B179$, 105 (1986).

\item{[10]}
P. Herczeg and R.N. Mohapatra, Phys. Rev. Lett. $\bf 69$, 2475 (1992).

\item{[11]}R.N. Mohapatra,  Phys. Rev. $\bf D34$, 909 (1986).

\item{[12]}P. Langacker, in $\it {Particle~ Data~ Tables}$, 1992;
see also, G. Altarelli, R. Barbieri and S. Jadach, CERN preprint
CERN-TH 6124/91.

 \item{[13]} CLEO collaboration, submitted to the Lepton-Photon Conference
at Cornell, August (1993).

\item{[14]}
see ${\it e.g.}$
A. Dar, J. Goodman and S. Nussinov, Phys. Rev. Lett. $\bf 58$, 2146 (1987).
F. Von Feilitzsch, in ${\it Neutrinos}$ ed. H. Klapdor, P.1
(Springer-Verlag, 1988).

\bye